\documentclass[sigconf,edbt]{acmart-edbt2020}
\usepackage{hyperref}
\usepackage{ulem}
\usepackage{lineno}
\usepackage{url}
\usepackage{microtype}
\usepackage{epsfig}
\usepackage{epstopdf}
\usepackage{subfigure}
\usepackage{amsmath}
\usepackage{comment}
\usepackage{array}
\usepackage{tikz}
\usepackage{pbox}
\usepackage{multirow}
\usepackage{enumitem}
\usepackage{todonotes}
\usepackage{caption}
\usepackage{mathtools}
\usepackage{breqn}
\usepackage{amsthm}
\usepackage{algorithm}
\usepackage[noend]{algpseudocode}

\algnewcommand\algorithmicforeach{\textbf{for each}}
\algdef{S}[FOR]{ForEach}[1]{\algorithmicforeach\ #1\ \algorithmicdo}

\begin{document}


\title{Detecting Stance in Tweets : A Signed Network based Approach}

\author{Roshni Chakraborty}
\affiliation{%
  \institution{Indian Institute of Technology Patna}
  \city{Patna}
  \country{India}}
\email{roshni.pcs15@iitp.ac.in}

\author{Maitry Bhavsar}
\affiliation{%
  \institution{Indian Institute of Technology Patna}
  \city{Patna}
  \country{India}}
\email{maitry.bhavsar@gmail.com}

\author{Sourav Kumar Dandapat}
\affiliation{%
 \institution{Indian Institute of Technology Patna}
  \country{India}}
\email{sourav@iitp.ac.in}

\author{Joydeep Chandra}
\affiliation{%
 \institution{Indian Institute of Technology Patna}
  \country{India}}
\email{joydeep@iitp.ac.in}




\begin{abstract}
Identifying user stance related to a political event has several applications, like determination of individual stance, shaping of public opinion, identifying popularity of government measures and many others. The huge volume of political discussions on social media platforms, like, Twitter, provide opportunities in developing automated mechanisms to identify individual stance and subsequently, scale to a large volume of users. However, issues like short text and huge variance in the vocabulary of the tweets make such exercise enormously difficult. Existing stance detection algorithms require either event specific training data or annotated twitter handles and therefore, are difficult to adapt to new events. In this paper, we propose a sign network based framework that use external information sources, like news articles to create a signed network of relevant entities with respect to a news event and subsequently use the same to detect stance of any tweet towards the event. Validation on $5,000$ tweets related to $10$ events indicates that the proposed approach can ensure over $6.5\%$ increase in average F1 score compared to the existing stance detection approaches. 
\end{abstract}


\maketitle

 
 \section{Introduction}
\label{intro}

\par A deeper understanding of the user opinions in Twitter can help in building automated feedback system that would be useful in applications, like government surveys~\cite{d2019monitoring}, product reviews and election opinion polls~\cite{lai2018stance}. Though several sentiment analysis techniques~\cite{mohammad2016sentiment,paradis2018visual}, that mainly classify a piece of text as positive or negative polarity, have been proposed, stance detection focuses on identifying the sentiment towards an  event from a text. For example, the tweet \textit{Corruption is sucking our country like leeches and has deep roots, what else other than demonetization could have worked} reflects a negative sentiment but has a positive stance towards the event \textit{Demonetization in India}\footnote{https://en.wikipedia.org/wiki/2016\_Indian\_banknote\_demonetisation}. Thus, the opinion towards the event may be obfuscated with extraneous information that makes stance detection a challenging task. Further, events are often represented through multiple targets\footnote{A \textit{target} is a mentioned entity (person, name) or a phrase in the text related to an event, towards which the author of the text expresses either a positive or negative sentiment.~\cite{krejzl2017stance}}
that are either used to express positive or negative sentiment towards the event. For example, sudden demonetization of certain currency notes in India in 2016  led to \textit{long ATM queues} but was expected to increase the \textit{economic progress} of the country. Both of these italicized phrases became representative target keywords of the event and were frequently used to express sentiment with respect to the event. Certain tweets supporting \textit{Demonetization in India} specifically mentioned positively about the possible economic growth, whereas those against it made negative mentions about the long queues and sufferings of the people. Thus, stance detection in tweets towards an event initially requires identification of the possible set of representative targets related to the event and the underlying polarity between the targets and the event and is different from identifying user stance towards an event. Further, the usage of non-vocabulary words, inherent noise, abbreviations, sarcasm~\cite{tsagkias2011linking} and absence of contextual information in tweets makes it a challenging job to detect stance from tweets.


\par To handle these issues, existing research works on stance detection in tweets rely on supervised or weakly supervised based approaches. Existing supervised approaches~\cite{shenoy2017performing,dey2017twitter} often rely on manually curated features, like target specific bag of words, event associated sentiments from the training set to determine stance of the tweets. However, these approaches require event-specific training data and predefined knowledge of the targets and thus, fail to provide a generic framework that can be adapted to newer set of events. Though existing weakly supervised algorithms reduce this dependency by using contextual information extracted from twitter handles of political personalities and hashtags, their accuracy diminishes when the pre-defined targets are not mentioned in the tweet texts~\cite{du2017stance}. Therefore, recent works~\cite{du2017stance} incorporate target specific embedding information to capture the stance related to a target with better accuracy. These existing approaches mostly consider predefined set of targets whose polarity towards an event is known apriori. However, such approach is not suitable for events where number of targets is huge and manual identification is difficult. Therefore, we intend to develop an automated stance detection system which can identify the targets automatically. However, implementing such a system would require addressing several challenges, such as identification of the targets of the event and automatic determination of polarity of those targets towards the event. We discovered recent works that consider the presence of multiple targets, however these works consider a very small set of specific targets, like \textit{Hillary Clinton}, \textit{Donald Trump}, \textit{Ted Cruz} and \textit{Bernie Sanders} whose political alignment are already known (e.g., \textit{Hillary Clinton} and \textit{Donald Trump} are always likely to oppose each other irrespective of the event)~\cite{sobhani2017dataset,wei2018multi} or require target specific training for deep learning based frameworks. However, it is highly difficult to attain target specific annotated dataset when the number of targets increase. Thus, we need to develop an automated stance detection technique which does not require target specific annotated dataset and can address these challenges.

\par In this paper, we propose a generic framework for automated stance detection that exploits extraneous information sources, like news articles, to identify the possible targets of the event. We create a signed network of the targets associated with the event that can be effectively used to determine the polarity relationship with the event. The novelty of our proposed approach is in identifying the relevant targets from the articles and generating the signed network representing the polarity of these targets towards the event or among themselves. The signed network representation of an event provides a holistic view of the polarity relationships of a set of targets towards the event which aids in identifying the stance of the tweet. Further, for those set of targets whose polarity relationship towards the event is not directly evident, the signed network representation provides a way to capture the same through application of network theoretic concepts. Identifying the stance of a tweet towards the most closely mapped targets in the signed network helps in identifying the stance expressed through the tweet. We rigorously validate our proposed approach  on $5,000$ tweets related to $10$ different events. Our investigations reveal that the proposed approach can ensure over $6.5\%$ increase in average F1 score than the existing approaches. 
The organization of the paper is as follows: We discuss few preliminaries related to the proposed approach in Section~\ref{s:prelim} followed by the related works in Section~\ref{s:rworks}. In Section~\ref{s:probF}, we present a formal definition of the problem followed by our proposed approach in Section~\ref{s:prop}. We discuss the experimental settings in Section~\ref{s:expt} and the observations in Section~\ref{s:results}. We finally draw our conclusions in Section~\ref{s:concl}.

\section{Definitions}\label{s:prelim}
In this Section, we describe some of the terminologies that have been frequently used in the paper. As these terminologies are already discussed in existing works, we highlight them for the ease of the readers. 

\begin{itemize}
    \item \textit{Target} : A \textit{target} is a mentioned entity (person, name) or a phrase in the text related to an event, towards which the author of the text expresses either a positive or negative sentiment~\cite{krejzl2017stance}. We refer to the phrases that are used as targets as \textit{key-phrases} and the entity as \textit{key-players} in this paper. 
    \item \textit{Polarity Relationship} : We consider the \textit{polarity} of a target towards an event as positive if it is used to express a positive  opinion about the event, and negative otherwise. The polarity between two targets is assumed to be positive if they are used to reflect a similar opinion about the event.  For example, the target phrase \textit{''crackdown on black money hoarders''} is considered as a positive effect of demonetization and hence would have a positive polarity towards the event, whereas target key-phrase like \textit{long ATM queues} would have a negative polarity. The polarity of the relationship between these key-phrases would be negative as they are used to reflect two different opinions about the event. The polarity of the relationship of a key-player towards an event would be positive if the key-player supports the event and negative otherwise.
    \item \textit{Signed Network} : It is a network in which each edge either has a positive or a negative sign~\cite{leskovec2010signed}, reflecting the polarity of the relationship among the connected nodes.
    \item \textit{Structural Balance of a Network} : A signed network is said to be structurally balanced if it can be divided into two groups, where the edges connecting nodes of a same group are always positive and those connecting nodes in different groups are negative~\cite{cartwright1956structural}.
\end{itemize}

\section{Related Works}\label{s:rworks}
There is a plethora of research works that determine stance from long texts like debates and discussions through feature based machine learning as well as deep learning techniques~\cite{mohtarami2018automatic,rajendran2018stance,zhang2018ranking,ruder2018360}. Existing feature driven research works consider textual features ~\cite{rajendran2018something}, contextual features~\cite{anand2011cats,hasan2013stance}, author's opinion history and interaction patterns~\cite{hasan2012predicting,trabelsi2018unsupervised}, deep learning models~\cite{mohtarami2018automatic}. However applying these methods directly to detect stance from social media contents, like Twitter, is difficult due to the absence of contextual information, presence of noise and usage of informal language~\cite{tsagkias2011linking}. Thus, we discuss the major works detecting stance on tweets related to different methodology next.

\subsection{Supervised Approach}  
Existing supervised techniques have studied the use of both feature driven machine learning algorithms~\cite{mourad2018stance,siddiqua2018stance} and deep learning models~\cite{wei2018multi,ma2018detect} to detect stance from short length texts. Existing feature based machine learning algorithms consider several features like the language components~\cite{skeppstedt2016active,simaki2018detection}, content attributes(n-grams, hashtags and named entities)~\cite{kuccuk2018stance,wojatzki2016ltl}, interactions between sentiment-stance-input variables~\cite{ebrahimi2016joint},
word co-occurrences~\cite{chuang2015stance}, presence of claims~\cite{bar2017stance}, contextual information~\cite{sasaki2016stance,lai2016friends,wojatzki2016ltl}, user previous information~\cite{sasaki2018predicting}, and target specific bag of keywords~\cite{shenoy2017performing,patra2016ju_nlp}. Creating target specific keywords  is widely used in stance detection and different mechanisms have been proposed to generate the same. While Shenoy et al.~\cite{shenoy2017performing} create a target specific bag of POS-tagged keywords along with sentiments, Patra et al.~\cite{patra2016ju_nlp} create target-specific topic bags that include dependency information among lexicons and hashtags to detect stance in tweets. Various machine learning frameworks, like an ensemble of classifiers~\cite{liu2016iucl} and character level and word level CNN~\cite{vijayaraghavan2016deepstance}  have been proposed. However, most of these existing approaches fail to detect stance related to multiple targets. Recent deep learning approaches have proposed attention mechanisms~\cite{wei2018multi,du2017stance,gao2018stance} or ensemble of several neural models~\cite{sobhani2017dataset,sobhaniexploring} identify stance related to multiple targets. However, these approaches consider a very small predefined set of targets who shares similar polarity relationship irrespective of an event, like \textit{Hillary Clinton}, \textit{Donald Trump} and \textit{Bernie
Sanders}, thus making it inapplicable for stance detection of tweets towards an event which might comprise of a large number of possible targets and further, the polarity relationship between the targets related to an event varies across events (contrary to the assumption of the existing works).
\par Therefore, stance detection approaches that rely on supervised algorithms are highly dependent on event-specific labeled data and hence, fail to provide a generic framework to detect stance in tweets. Further, these approaches utilize a set of event related tweets or a predefined list of targets related to an event which cannot cover the entire range of possible targets related to the event or their polarity relationships and thus, fail to ensure stance detection of tweets related to an event. Our empirical analysis indicates that there could be a large number of possible targets related to an event (as shown in table~\ref{tab:TarStat}) which is difficult to be captured by the existing supervised research works on stance detection of a tweet. Further, most of the existing works don't consider capturing the polarity of the target towards an event which is of high importance. 

\subsection{Weak Supervised Approach}

Existing supervised learning approaches require huge annotated data related to each event for effective stance detection in tweets. In order to reduce this dependency on labeled dataset, there is a line of research works that have proposed machine learning frameworks that require weak supervision. Weak supervised stance detection works rely on different characteristics related to an event, like stance related information from annotated hashtags~\cite{dias2016inf,misra2017semi,krejzl2016uwb}, user behavior and information~\cite{klenner2017stance,rajadesingan2014identifying,fraisier2018stance}, specific key-phrases related to events~\cite{johnson2016identifying} and topical information~\cite{dey2018topical}. However, all of these approaches require annotated event-specific twitter handles, hashtags or key-phrases. Further, as these proposed approaches are largely dependent on the annotated hashtags and twitter handles, it can't ensure complete coverage of all the related targets towards the event. Further, most of these approaches rely on event-specific keywords, relationships among targets and sentiment dictionary that might require significant manual annotation.
\par Several deep learning based frameworks which require less amount of labelled information than the supervised deep learning approaches has been proposed, like convolutional neural network~\cite{wei2016pkudblab}  that utilizes word embedding from Google News database and bidirectional LSTM model~\cite{augenstein2016stance} utilizes stance related to a target to determine stance related to different targets. Although this approach can detect stance from tweets even when the target name is not explicitly mentioned, it is applicable only for targets for whom the polarity is already known, like \textit{Hillary Clinton} and \textit{Donald Trump} who mostly belong to different polarity irrespective of the event. However, an event might have a large set of possible targets and their polarity among themselves and also with the event might differ across the events and hence, knowledge about very specific targets can't ensure stance detection in tweets related to any event. Further, the deep learning frameworks do require event specific training data which makes it dependent on user intervention and  event specific knowledge for it's applications. 
\par Thus, none of the existing research works that detect stance from tweets provide a generic mechanism which does not require event specific labeled dataset. Information extracted from annotated twitter handles, labeled training set or stance specific hashtags can not cater to the huge vocabulary diversity related to an event. Further, most of the existing works do not consider the polarity relationship between the targets or the target with the event which makes it inapplicable to detect stance of a tweet related to an event. Although few recent research works incorporate an understanding of the polarity between multiple targets, they consider a very small set of pre-defined targets, like \textit{Hillary Clinton} and \textit{Donald Trump} making it inapplicable for stance detection in tweets towards an event which comprises of a huge set of possible targets and different polarity relationships. Therefore, we propose a generic framework  that can automatically identify the possible set of targets related to an event and the underlying relationship (positive or negative) of those targets towards the event. Extraction of this knowledge and representation through a signed network ensures detection of stance in tweets with high accuracy irrespective of the inherent limitations of tweets, like absence of contextual information and presence of inherent noise in tweets. We discuss the proposed approach in details in section~\ref{s:prop} and provide a formal definition of the problem along with an overview of the proposed approach next.

\begin{table}
\centering
\begin{tabular}{|c|c|c|c|c|c|c|}
      \hline
      \textbf{Event} & \textbf{$K_{p_Tw}$} & \textbf{$K_{l_Tw}$} & \textbf{$Tw$} & \textbf{$K_p$} & \textbf{$K_l$} & \textbf{$Nw_{st}$}\\
      \hline
      {$E_{1}$} & 28 &  18 & 5000 & 37 &  110 & 612\\ 
      \hline
      {$E_{2}$}   & 22 & 35 & 5000 & 102 & 112 & 1720\\
      \hline
      {$E_{3}$} & 53  & 27 & 5000   & 73  & 34 & 987\\
      \hline
\end{tabular}
\caption{The number of targets extracted from \textit{tweets}($Tw$) and \textit{news articles}($Nw$) related to each of the $10$ events is shown. The target set related to an event extracted from \textit{news article sentences}($Nw_{st}$) are key-phrases as $K_p$ and key-players as $K_l$ respectively and those from \textit{tweets} are key-phrases as $K_{p_Tw}$ and key-players as $K_{l_Tw}$.}\label{tab:resStat}
\end{table}
\section{Overview of the Problem}\label{s:probF} 
Assuming an event $E$ and a set of tweets $Tw$ related to $E$, the objective is to find the stance (positive, negative or neutral) of each tweet ($Tw_z$) in $Tw$ with respect to $E$. Therefore, we assume that the event $E$ is associated with unknown targets, $g_1,g_2,\ldots,g_m$, each of which has certain polarity (positive or negative) towards $E$ and $Tw_z$ mentions a subset of these targets. If $\mathcal{S}(Tw_z,g_i)$ denotes the sentiment  (positive, negative or neutral) of the tweet $Tw_z$ towards $g_i$, and $\mathcal{R}(E,g_i)$ represents the polarity of the relationship of $g_i$ towards the event $E$, then the stance, $\mathcal{X}(Tw_z,E)$ of $Tw_z$ towards $E$ would be represented as 
\begin{eqnarray}
\mathcal{X}(Tw_z,E)=\mathcal{S}(Tw_z,g_i)\mathcal{R}(E,g_i)\label{eq:stanceEq}
\end{eqnarray}
As evident from Equation~\ref{eq:stanceEq}, the stance of a tweet is positive if either the tweet reports positively  about a target that is positively related to the event (both $\mathcal{S}(Tw_z,g_i)$ and  $\mathcal{R}(E,g_i)= positive$), or it reports negatively about a target that in turn is negatively related to $E$ (both $\mathcal{S}(Tw_z,g_i)$ and $\mathcal{R}(E,g_i)=negative$).
Thus, our stance detection approach consists of three major sub-tasks:
\begin{enumerate}
\item Identification of the targets of the event.
\item Identification of the polarity relationship among the targets and the event.
\item Identification of the sentiment expressed through a tweet towards the target.
\end{enumerate}
\par
\begin{figure*}[ht]
\centerline{
\includegraphics[width=6in]{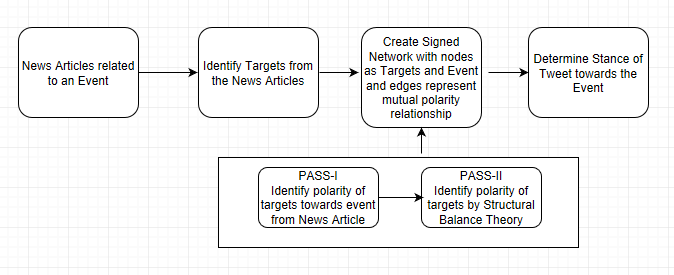}}
\caption{Block Diagram Representing the Proposed Approach}
\label{fig:name}
\end{figure*}

 As already discussed, the targets related to an event can either be a phrase or an entity and there can be a number of possible targets related to the event. Identifying the targets directly from the tweets requires automatic determination of the relevance of a target towards an event from the tweet text. For that, we require manual labeling the appropriate targets which is extremely tedious and requires a huge volume of tweets to capture all the relevant targets. To validate this claim, we conducted a study in which we simultaneously applied our proposed target extraction approach (details in Section~\ref{s:tarEx}) on $5000$ tweets and news articles related to an event. We considered $3$ different events for this study. Our observations as shown in Table~\ref{tab:resStat} indicate a small fraction of the targets could be identified from the tweets as compared to the news articles. To investigate whether this difference can be reduced by increasing the size of the tweet set, we increased the tweet set size to $22,000$ for each of the $3$ events. We did not observe any significant change in the number of targets identified. These observations motivated the use of external information sources, like, news articles to identify the targets. We intuitively believe as the sentences in a news articles are more focused towards the event than the tweets, it helps in precise identification of the targets. We propose a a signed network based approach to determine the polarity relationship of the targets which could be eventually to identify the stance of a tweet. An outline of the steps of the proposed approach is provided in Figure~\ref{fig:name} and we describe each of the steps in details next. 

\section{Proposed Approach}\label{s:prop}
In this Section, we discuss the proposed approach in details.

\begin{table*}[ht]
\centering
\begin{tabular}{|c|c|c|c|}
      \hline
      \textbf{SNo} & \textbf{Tweet} & \textbf{Target} & \textbf{Stance}\\
      \hline
      {1}  & \pbox{9cm}{This is a major step by BJP government to remove black money } &  BJP government & Positive\\ 
      \hline
      {2}  & \pbox{9cm}{cashless transactions has reduced Terrorism} &  cashless transactions & Positive\\ 
      \hline
      {4}   & \pbox{9cm}{note ban has caused huge collateral damage to the Indian economy} & note ban & Negative\\
      \hline
      {5}  & \pbox{9cm}{ Mamata claims Modi is looting people's money by note ban} &  Mamata & Negative\\ 
      \hline
\end{tabular}
\caption{Tweets that indicate stance related to the event \textit{Demonetization in India} but does not comprise of the event name in the tweet are shown.} \label{tab:TarData}
\end{table*}

\subsection{Identification of Targets of the Event} \label{s:tarEx}
Identifying the targets with respect to a news event is important for stance detection. There are several challenges, like high diversity of the keywords in the tweets and presence of a large number of possible targets for an event and the targets mentioned in the tweets can either be key-phrases or key-players related to the event. In Table~\ref{tab:TarData}, we show few example tweets related to the \textit{Demonetization} event. We discuss how we automatically extract the key-phrases and the key-players with respect to an event next.

\par \textit{Extraction of key-phrases:} We use RAKE~\cite{rose2010automatic}, a key-phrase technique that computes the importance score of the phrases based on the importance of the constituent words with respect to the event and their co-occurrence frequency. If $w_1$, $w_2$,\ldots $w_m$ be the constituent words in the phrase $P_p$, then the phrase importance score, $I(P_p)$ would be given as 


 \begin{eqnarray}
 I(P_p) &= f(P_p)*\sum_{i=1}^{m} g(w_i)
 \end{eqnarray}
where, $f(P_p)$ represents the frequency of co-occurrence of the words of $P_p$ in the news article and $g(w_i)$ represents the frequency of the word $w_i$ in the news article. 
We select the phrases which have an importance score higher than the threshold $\theta_p$ (i.e., $I(P_p)$ $\geq$ $\theta_p$) as the \textit{key-phrases}, $K_p$ of the event. The choice of the threshold is determined based on a manual study by $3$ different annotators on around $1270$ phrases selected from news articles of $3$ different events. We asked the annotators to identify those phrases that could be considered as key-phrases, without looking at their importance scores. We observed that the importance scores that are one standard deviation greater than the mean can be considered as target key-phrases. In Table~\ref{tab:KPStat}, we show the number of phrases extracted by RAKE, the number of key-phrases which have an importance score higher than the threshold along with the maximum, mean and the standard deviation of the importance score of the phrases (details of the events are given in Section~\ref{s:dataset}). We show the key-phrases through word cloud in Figure~\ref{fig:rake_1}.
\begin{table}[ht]
\centering
\begin{tabular}{|c|c|c|c|c|c|c|c|}
      \hline
      \textbf{Event} & \textbf{$p$} & \textbf{$K_p$} & \textbf{Max $I(P_p)$} & \textbf{Mean $I(P_p)$} & \textbf{SD $I(P_p)$}\\
      \hline
      {$E_{1}$} & 147 &  37 & 7.04 & 2.13  & 1.08\\ 
      \hline
      {$E_{2}$}   & 702 & 102 & 12.50 & 2.41 &  1.68\\
      \hline
      {$E_{3}$} & 417  & 73 & 12.29   & 2.23 & 1.45\\
      \hline
\end{tabular}
\caption{The number of phrases extracted by RAKE, the number of key-phrases which have an importance score higher than the threshold, maximum importance score, average importance score and the standard deviation of the importance score of the phrases related to $3$ events is shown.}\label{tab:KPStat}
\end{table}

\par \textit{Extraction of key-players:} We use SpaCY\footnote{honnibal.github.io/spaCy} tool to extract the names of all the persons who have featured in the news articles related to the event. The set $K_l$ comprises of all the collected names is the key-players set of the event. We form the possible set of targets of the event as the set of key-phrases and key-players identified related to an event. In Table~\ref{tab:TarStat}, we show the number of key-phrases ($K_p$) and key-players ($K_l$) identified for $10$ events (the details of the events is discussed in Section~\ref{s:dataset}). We next discuss the procedure to identify the polarity relationship of the targets towards the event. 


\begin{table}[ht]
\centering
\begin{tabular}{|c|c|c|c|c|c|c|c|}
      \hline
      \textbf{$E$} & \textbf{$K_p$} & \textbf{$K_l$} & \textbf{$E$} & \textbf{$K_p$} & \textbf{$K_l$}\\
      \hline
      {$E_{1}$} & 37 &  110 & {$E_{6}$} & 58  & 28\\ 
      \hline
      {$E_{2}$}   & 102 & 112 & {$E_{7}$} & 27 &  37\\
      \hline
      {$E_{3}$} & 73  & 34 & {$E_{8}$}   & 82 & 63\\
      \hline
      {$E_{4}$} & 20 &  32 & {$E_{9}$} & 46  & 68\\ 
      \hline
      {$E_{5}$}   & 67 & 33 & {$E_{10}$} & 73  & 64\\
      \hline
\end{tabular}
\caption{The number of targets, i.e., $K_p$ and $K_l$, related to each of the $10$ events is shown.}\label{tab:TarStat}
\end{table}


\subsection{Identification of polarity of a target towards the event}\label{s:sgnW}

We adopt a two-pass method to extract the relationships among targets. In Pass-I, we utilize the news articles to extract either $a)$ directly the polarity of a target towards the event or, $b)$ the polarity of a target towards another target. Due to the inherent complexity of the sentences, it is very challenging to capture the polarity relationships of all the targets toward the event. Therefore, in Pass-II, we create a signed network of the targets from the relations derived in Pass-I and then, determine the polarity relationship of the unresolved targets by structural balance theory~\cite{cartwright1956structural,derr2018signed}. Therefore, the proposed approach integrates the knowledge from the textual contents and signed network based concepts to identify the polarity of targets towards an event. We discuss in details each of these steps next.

\subsubsection{Pass-I: Identifying target polarity based on sentence types}\label{s:newsArt}
In a sentence, the  polarity relationships can be identified depending on the type of speech (direct or indirect speech) and the sentence type (simple, compound or complex). For example, in a simple sentence like \textit{{INC party to begin nationwide campaign against demonetization}}, the polarity relation between target \textit{INC} and the event \textit{demonetization} could be directly established by lexicon based sentiment analysis by Taboada et al.~\cite{taboada2011lexicon}. However, for complex sentences,  requires understanding of the conjunction and the clauses present in the sentence to extract the polarity relationship~\cite{savanur2017feature}. Further, extraction of polarity relationships from direct or indirect speech requires both identification of the speaker and his view expressed through the speech.

\par  In \textit{Pass-I}, we initially identify sentences that are direct or indirect speeches by the presence of reporting verbs (an exhaustive list of reporting verbs is given by Krestel et al.~\cite{krestel2008minding}) and further, segregate the speech into either direct or indirect speech based on the presence of \textit{quotes} and finally, classify each sentence which are neither direct nor indirect speech into simple, complex or compound sentences based on the clauses and the conjunctions present in the sentence following the approach proposed by Puscaso et al.~\cite{puscasu2004multilingual}. 

\par \textbf{Simple Sentences}: As a simple sentence has only one main clause, we directly use the lexicon based approach proposed in Taboada et al.~\cite{taboada2011lexicon} on the extracted clause to determine the polarity of a target towards the event or towards another target present in the clause by capturing the polarity of the sentence towards the subject. We calculate the polarity of the sentence by the semantic orientation which considers both polarity and strength of each lexicon of the sentence. The semantic orientation of each lexicon is calculated on the basis of \textit{Semantic Orientation
CALculator} (SO-CAL) \footnote{http://www.cs.sfu.ca/~sentimen/socal/login.php?page=111} that considers both the strength and polarity of each lexicon,  presence of intensifiers in the neighborhood, like \textit{most} and \textit{slightly} either increase or decrease, respectively, the semantic orientation of the lexicon and the presence of negations reverse the polarity of the lexicon. Thus, if $w_1, w_2\ldots, w_n$ represent the lexicons in a sentence, $s$, and have semantic orientations as $S(w_1), S(w_2),\ldots, S(w_n)$ respectively, then the semantic orientation, $S(s)$, is given as
     \begin{eqnarray}
        S(s) &= \sum_{i=1}^{n} S(w_i)\label{eq:SO}
    \end{eqnarray}
$S(w_{i+1})$ is also affected by the presence of intensifier or negation as $w_i$.  If $w_i$ is an intensifier or negation, then the following rule applies for calculating $S(w_{i+1})$: 
\par \textit{Itensifier} : An intensifier can either be an amplifier or downtoner. When $w_i$ is an amplifier with an amplifier score of $I(w_i)$, then the semantic orientation of $w_{i+1}$ is increased by  $S(w_{i+1})\frac{I(w_i)}{100}$ and would be given as $S_{w+1}(1+\frac{I(w_i)}{100})$. For example, if $w_i$ is the lexicon \textit{very} which is an amplifier and $I(very)$  is $+25$ and the lexicon(i.e., $w_{i+1}$) is \textit{good} with $S(good)$ being $+3$, then $S(very$ $good)$ is calculated  as $S(good) * (1 + I(very)/100)$ which is $3 * (1 + 0.25)$, i.e., $+3.75$. Similarly, when $w_i$ is a downtoner, then then the semantic orientation of $w_{i+1}$ is decreased by $S(w_{i+1})\frac{I(w_i)}{100}$.

\par \textit{Downtoner} :When $w_i$ is a negation, then the semantic orientation of $w_{i+1}$ is decreased by the semantic orientation of $w_i$, and would be given as $S(w_{i+1})-S(w_i)$. For example, if $w_i$ is the lexicon \textit{not} which is a negation with $S(not)=4$ followed by the lexicon \textit{good} (i.e., $w_{i+1}$) with $S(good)$ being $+3$, then $S(not$ $good)$ is calculated  as ${S(good) - S(not)}$ which is $(3 - 4)$, i.e., $-1$.
 To calculate the semantic orientation of the sentence $s$, we calculate the semantic orientation of the constituent lexicons $w_i$ and apply equation \ref{eq:SO} to derive the semantic orientation of the sentence, $S(s)$ that represents the polarity relationship between the subject and the object in $s$. Therefore, if ($S(s) > 0$), then the polarity between the subject and the object is taken as positive and if ($S(s) < 0$), then the polarity between the subject and the object is taken as negative. The subject and the object represents either a pair of targets or a target and the event respectively. 

\par \textbf{Compound and Complex Sentence}: 
Determination of the polarity relationship from a compound or complex sentence requires understanding of the present clause and conjunction. We rely on the approach proposed by Puscasu et al. ~\cite{puscasu2004multilingual} to identify the clauses, identify the sentiment of each clause by ~\cite{taboada2011lexicon}. We, finally, follow Savanur et al.~\cite{savanur2017feature} to determine the clause boundaries and then, identify the polarity relationship of the sentence on the basis of the rules regarding each conjunction. We summarize the rules~\cite{savanur2017feature} that we follow in Table~\ref{tab:ConjData}. Although there can be other conjunctions that are not mentioned in the table~\ref{tab:ConjData}, we consider only these mentioned conjunctions. For example, complex sentences that express \textit{cause-effect relations}, like sentences with the subordinating conjunction, \textit{because} requires understanding of only the \textit{effect} relation (by rule $1$ in table~\ref{tab:ConjData}).

\par \textbf{Direct and Indirect Speech}: For sentences that express direct or indirect speech, we identify the speaker and the polarity of the speaker towards the subject mentioned in the \textit{reporting clause}. We identify the speaker from direct or indirect speech by Krestel et al.~\cite{krestel2008minding} and use Taboada et al.~\cite{taboada2011lexicon} to calculate the polarity. Although the \textit{reporting clause} might be a complex or compound sentence, the proposed approach considers only those direct and indirect speech in which the \textit{reporting clause} is a simple sentence.

   \begin{table*}[ht]
\begin{tabular}{|c|c|c|c|}
      \hline
      \textbf{No} & \textbf{Conjunction} & \textbf{Sentence} & \textbf{Rule}\\
      \hline
      {1}  & \pbox{3cm}{cause and effect relation shown by \textit{because, as, since}} &  \pbox{6cm}{Bihar CM supported Demonetization because he believed it could curb corruption.}& \pbox{3cm}{consider the effect relation}\\ 
      \hline
      {2}  & \pbox{3cm}{\textit{although, though}} &  \pbox{6cm}{Although the opposition parties have different ideologies, they are united against Prime Minister Modi.}& \pbox{3cm}{consider the subordinate clause}\\ 
      \hline
      {3}  & \pbox{3cm}{\textit{but, yet}} &  \pbox{6cm}{All opposition parties may march separately, but they have attacked the demonetization move together}& \pbox{3cm}{consider the clause following \textit{but} or \textit{yet} conjunction}\\ 
      \hline
      {4}  & \pbox{3cm}{\textit{and, or}} &  \pbox{6cm}{Demonetization has caused political disappointment and people are frustrated waiting in the long queues at ATM}& \pbox{3cm}{consider both the clauses}\\ 
      \hline
      {5}  & \pbox{3cm}{relative clauses shown by \textit{who, which, whom, that, whose}} &  \pbox{6cm}{People in villages don't trust cashless transactions which Modi is advertising for.}& \pbox{3cm}{consider the main clause}\\ 
      \hline
\end{tabular}
\caption{The rules related to different types of conjunctions as discussed in existing research works~\cite{savanur2017feature} and \cite{puscasu2004multilingual} are shown here.}\label{tab:ConjData}
\end{table*}

Our observations on $10$ events (with article count varying from $10$ to $86$ for each event) reveal that the percentage of targets whose polarity relationship towards the event is resolved by \textit{Pass-I} varies from $14-62.5\%$ (as shown in Table~\ref{tab:missLink}). Therefore, we observe that we could resolve the polarity for a significant proportion of the targets which we resolve by creating a signed network of the targets and the event.


\subsubsection{Pass-II: Identifying target polarity using signed network analysis}\label{s:missL}
We propose utilization of \textit{structural balance theory}~\cite{cartwright1956structural} in signed networks to resolve the polarity of those targets which could not be resolved in Pass-I towards the event. Structural balance theory in sentiment networks~\cite{rawlingSociology2017} shows that these networks move towards specific structural configurations over time. The structural balance theory~\cite{burt2000network,leskovec2010signed,cartwright1956structural} states that the structural balance of a triad micro-structure in the network is determined by the configurations of the positive and negative relations among the nodes in the triad where positive relations signify \textit{friends} and negative relations as \textit{enemies}. Further, a sentiment network is said to be structurally balanced if it holds all the following properties:
\begin{enumerate}
    \item ($P_1$) A friend of a friend is friend
    \item ($P_2$) A friend of an enemy is enemy
    \item ($P_3$) An enemy of a friend is an enemy
    \item ($P_4$) An enemy of an enemy is a friend
\end{enumerate}
Each of the properties of structural balance triads can be directly applied to determine the polarity relationship of certain targets, whose individual polarity relationships with the event are not known. For example, if a target $g_i$ shares a positive polarity with another target $g_j$ whose polarity relationship towards the event is known, then based on properties $P_1$ and $P_3$ of the balance model, $g_i$ would share the same polarity relationship with $E$ as $g_j$. On the other hand if $g_i$ shares a negative polarity with $g_j$, then based on properties $P_2$ and $P_4$ its polarity relationship towards $E$ would be opposite to that of $g_j$. This is quite intuitive in the sense that a positive polarity between targets $g_i$ and $g_j$ implies that they are used to convey the same sentiment towards the event, hence the polarity relationship of $g_i$ towards $E$ would be the same as $g_j$ itself. On the other hand, a negative polarity between targets $g_i$ and $g_j$ indicates that they are used to convey different opinions and hence, $g_i$ must share a different polarity towards $E$ as $g_j$. We use the signed network to derive the polarity relation $\mathcal{R}(E,g_i)$ of a target $g_i$ towards the event, $E$, for cases where the same could not be derived directly from Pass-I, if $g_i$ satisfies one of the following cases:

\begin{enumerate}
    \item Target $g_i$ is not directly connected to $E$ but is connected to another target $g_j$ 
    that in turn is connected to $E$ 
    .
    \item Target $g_i$ is connected to $E$ through a path $g_i, g_{i+1}, g_{i+2},\ldots, g_{i+k}$, where the polarity of relationship between targets $g_i$ and $g_{i+1}$ is known. 
\end{enumerate}

We maintain the properties of structural balance in the target network and use the following construct rules to derive the polarity relationships of the target $g_i$ towards the event.\\

\textit{Case 1} : Based on properties $P_1$ to $P_4$, the polarity $\mathcal{R}(E, g_i)$ of $g_i$ towards $E$ would be given as 
    \begin{eqnarray}
    \mathcal{R}(E, g_i)&=&\mathcal{R}(g_i, g_j)\mathcal{R}(E,g_j)\label{eq:R-direct}
    \end{eqnarray}
    \textit{Case 2} : To determine the polarity of relation of target $g_i$ towards $E$, we use a method proposed in~\cite{kim2018side}. We add a hypothetical edge from each of the targets $g_i$, $g_{i+1}$, \ldots, $g_{i+k-1}$ to event $E$. Thus, again based on the properties $P_1$ to $P_4$, the polarity of relationship of $g_i$ towards $E$ would then be given as 
 \begin{eqnarray}
 \mathcal{R}(E,g_i)&=&\mathcal{R}(g_i, g_{i+1})\mathcal{R}(E, g_{i+1})\nonumber\\
 &=&\mathcal{R}(g_i, g_{i+1})\mathcal{R}(g_{i+1}, g_{i+2})\mathcal{R}(E, g_{i+2})\nonumber\\
 &=&\mathcal{R}(g_i, g_{i+1})\cdots\mathcal{R}(g_{i+k-1}, g_{i+k})\mathcal{R}(E, g_{i+k})\label{eq:R-indirect}
 \end{eqnarray}
We next prove using a theorem that the polarity relationships derived using the above construct rules clusters the network into two groups of positively connected targets, where the polarity of all the targets of one group is positive towards $E$ and those in the other group have negative polarity towards $E$. This also implies that the polarity relation between a pair of targets in two different groups would always be negative.
\begin{theorem}
Following the construct rules of cases $1$ and $2$, if the target network is clustered into 2 groups, $G^{+}$ and $G^{-}$, based on the polarity relation $\mathcal{R}(E, g_i)$ of the nodes $g_i$ being $(+)ve$ or $(-)ve$, respectively, then if for any two target nodes $g_i$ and $g_j$, $\mathcal{R}(g_i,g_j)$ is $(+)$ve then $g_i,g_j\in G^{+}$ (or $G^{-}$).
\end{theorem}
\textit{Proof} 
If $\mathcal{R}(g_i,g_j)$ is $(+)$ve then based on equation~\ref{eq:R-direct}, $\mathcal{R}(g_i, g_j)=\frac{\mathcal{R}(E, g_i)}{\mathcal{R}(E, g_j)}>0$, thus indicating that both $\mathcal{R}(E, g_i)$ and $\mathcal{R}(E, g_j)$ must both be either positive or negative. Thus both $g_i$ and $g_j$ would either belong to $G^{+}$ or $G^{-}$.

\par The above theorem implies that by observing the mutual polarity relation between any pair of targets we can determine the polarity relation with respect to the event if it shares only either positive or negative polarity relationship with one or more targets in the same group. We investigate the effectiveness and accuracy of the proposed approach in Section~\ref{s:results}. We observe that nearly $88-94\%$ of the targets' polarity could be resolved by combining Pass-II with Pass-I as shown in Table~\ref{tab:missLink}. Therefore, our observations highlight the importance of unifying information from both the news articles and structural balance theory of signed networks to resolve the polarity of the targets of an event. We next discuss the final step whereby we extract the sentiment of the tweet towards one or more targets and use the polarity relationships of the targets to derive the stance of the tweets.

\begin{table}[ht]
\centering
\begin{tabular}{|c|c|c|c|c|}
      \hline
      \textbf{Event} & \textbf{Pass-I} & \textbf{Pass-II} & \textbf{Unresolved Links}\\
      \hline
      
      {$E_{1}$} & 0.527 &  0.428 & 0.04\\ 
      \hline
      {$E_{2}$}   & 0.14 & 0.76 & 0.10\\
      \hline
      {$E_{3}$} & 0.277  & 0.643 & 0.08\\
      \hline
      {$E_{4}$}   & 0.33 & 0.55 & 0.12\\
      \hline
      {$E_{5}$} & 0.625  & 0.313 & 0.06\\
      \hline
      {$E_{6}$} & 0.30  & 0.60 & 0.10\\
      \hline
      {$E_{7}$} & 0.303  & 0.576 & 0.12\\
      \hline
      {$E_{8}$}   & 0.351 & 0.598 & 0.051\\
      \hline
      {$E_{9}$} & 0.296 & 0.593 & 0.11\\
      \hline
      {$E_{10}$} & 0.40  & 0.52 & 0.08\\
      \hline
\end{tabular}
\caption{The fraction of targets' polarity relationship detected by analyzing news articles and structural balance theory respectively along with the fraction of unresolved links related to each of the $10$ events is shown.}\label{tab:missLink}
\end{table}
\subsection{Extraction of the sentiment of the tweet towards the event}\label{s:senTw}
The objective of this step is to extract the sentiment of the tweet towards the event.  
As the sentiment towards the event may not be directly evident from the tweet, we initially filter the target, $g_i$ mentioned in the tweet text ($Tw_z$) as either the event ($E$) or a target related to the event (from the list of targets, $T$) and measure the sentiment of the tweet text towards $g_i$. The calculation of the sentiment of the tweet text towards $g_i$, $\mathcal{S}(Tw_z,g_i)$ is done on the basis of lexicon based sentiment analysis approach~\cite{taboada2011lexicon}. We use the polarity relation $\mathcal{R}(E,g_i)$ of the target mentioned in the tweet text, $g_i$ towards the event, $E$ from the created signed network and consider both the relationships $\mathcal{S}(Tw_z,g_i)$ and $\mathcal{R}(E,g_i)$ to determine the stance of the tweet towards the event, $\mathcal{X}(Tw_z,E)$ by the relation.
\begin{eqnarray}
\mathcal{X}(Tw_z,E)=\mathcal{S}(Tw_z,g_i)\mathcal{R}(E,g_i)
\end{eqnarray}
Therefore, $\mathcal{X}(Tw_z,E)$ is positive if $\mathcal{R}(E,g_i)$ and $\mathcal{S}(Tw_z,g_i)$ are either both positive or negative. We discuss the dataset details and experimental settings next.

\section{Experimental Settings}\label{s:expt}
In this Section, we discuss the events that we consider for the experiments and outline the procedure to create the ground truth data, discuss the existing research works with which we compare the proposed approach.

\subsection{Dataset Collection and Pre-processing Details}\label{s:dataset}
For our experiments, we randomly select $10$  events that had occurred from $2016$ to $2018$, from the Wikipedia list of events\footnote{https://en.wikipedia.org/wiki/2018}\footnote{https://en.wikipedia.org/wiki/2017}. To avoid any bias towards events with more number of news articles, we selected events with varying number of news articles --- the total number of news articles related to an event ranged between $10-86$. We choose events with different number of news articles to understand the difference in performance of our proposed approach with respect to the number of articles available for an event. The total number of news articles related to an event was considered based on the news articles shown by Google News Search API~\footnote{https://news.google.com/?hl=en-IN\&gl=IN\&ceid=IN:en} related to that event. The details about the number of articles for each dataset along with the average number of sentences is outlined in Table~\ref{tab:newsArtData}. In Section~\ref{s:Nres}, we further investigate other characteristics of the news articles, like the average number of targets present in a news article along with the length of the news articles related to each of $10$ events to ensure the proposed approach is not biased by the selection of the news articles or the type of news event.  
\subsubsection{Dataset of the Events}\label{s:eve}
The following events were selected randomly to investigate the proposed approach.
\begin{itemize}
    \item \textbf{Harvey Weinstein allegations} ($E_{1}$) : In October $2017$, around $80$ women had made sexual abuse allegations against Harvey Weinstein, a former US film producer, that created a wave against sexual harrassment at work places\footnote{https://en.wikipedia.org/wiki/Harvey\_Weinstein}. 
    In Twitter, people expressed their opinion either in favour or against Harvey Weinstein.
    \item \textbf{Demonetization in India} ($E_{2}$) : On $8$ November $2016$, the Government of India announced the demonetisation of all $500$ and $1000$ Indian rupee banknotes of the Mahatma Gandhi Series \footnote{https://en.wikipedia.org/wiki/2016\_Indian\_banknote\_demonetisation}. People expressed their 
    support or criticized the decision of demonetization in India on Twitter.
    \item \textbf{Catalan Independence Movement} ($E_{3}$) : The Catalan independence movement is a social and political movement with roots in Catalan nationalism, which seeks the independence of Catalonia from Spain which led to a resolution that was passed by the Parliament of Catalonia on $27$ October $2017$, which declared the independence of Catalonia from Spain and the founding of an independent Catalan Republic\footnote{https://en.wikipedia.org/wiki/Catalan\_independence\_movement}. There was segregation in the opinion of people on this event as people either supported or opposed the Catalan Independence movement.
    
       \item \textbf{Australian Ball Tampering Scandal} ($E_{4}$) : In March 2018, the Australian cricket team was involved in a ball-tampering scandal during and after the third Test match against South Africa in Cape Town \footnote{https://en.wikipedia.org/wiki/2018\_Australian\_ball-tampering\_scandal}. While there was a section of people who deeply criticized the three Australian players involved and demanded huge punishment, another section of people were sympathetic and believed the punishment meted was harsh and the players should be forgiven.
       
        \item \textbf{Punjab National Bank Scam} ($E_{5}$) : The Punjab National Bank, a nationalized bank of India, made headlines after a case of fraudulent letter of undertaking worth $14,356.84$ crore rupees issued by the Punjab National Bank surfaced and the person responsible, \textit{Nirav Modi} and his family absconded the country\footnote{https://en.wikipedia.org/wiki/Punjab\_National\_Bank\_Scam}. It led to discussions about whether Indian government is negligent and responsible in allowing the accused to flee the country.
        
        \item \textbf{Assassination of Jamal Khashoggi} ($E_{6}$) : Jamal Ahmad Khashoggi was a Saudi Arabian dissident, author, columnist for the Washington Post, and a general manager and editor-in-chief of Al-Arab News Channel who was allegedly assassinated at the Saudi Arabian consulate in Istanbul on $2$ October $2018$\footnote{https://en.wikipedia.org/wiki/Jamal\_Khashoggi}. Khashoggi was primarily declared as \textit{missing} and Saudi Arabia claimed he had left the consulate and denied having any knowledge about his fate which led to debate among people that whether US government should interfere and ensure justice for Khashoggi.
        
        \item \textbf{2017 Zimbabwean coup d'état} ($E_{7}$) : On the evening of 14 November 2017, elements of the Zimbabwe Defence Forces (ZDF) gathered around Harare, the capital of Zimbabwe, and seized control of the Zimbabwe Broadcasting Corporation and key areas of the city which led to removal of President Mugabe succeeded by Emmerson Mnangagwa\footnote{https://en.wikipedia.org/wiki/2017\_Zimbabwean\_coup\_d\%27\%C3\%A9tat}.
         \item \textbf{United States withdrawal from the Joint Comprehensive Plan of Action} ($E_{8}$) : On May 8, 2018, the United States withdrew from the Joint Comprehensive Plan of Action (unofficially known as the \textit{Iran Deal} or the \textit{Iran Nuclear Deal}). The withdrawal caused concerns in Iran due to its impact on the economy and received mixed reactions, like praise from the American conservatives in the United States and criticism from the former president Barack Obama, his vice president Joe Biden and EU\footnote{https://en.wikipedia.org/wiki/United\_States\_withdrawal\_from\_the\_Joint\_Comprehensive\_Plan\_of\_Action}.
        
        \item \textbf{2017 Nangarhar airstrike} ($E_{9}$) : The 2017 Nangarhar airstrike refers to the American bombing of the Achin District located in the Nangarhar Province of eastern Afghanistan, near the border with Pakistan with the goal of destroying tunnel complexes used by a branch of the Syria-based ISIS\footnote{https://en.wikipedia.org/wiki/2017\_Nangarhar\_airstrike}.
        
        \item \textbf{Otto Warmbier} ($E_{10}$) : Otto Frederick Warmbier was an American college student who was imprisoned in North Korea in 2016 after being convicted of theft of a propaganda poster. Although the U.S. government secured his release in June 2017, Warmbier died six days after his return to the United States. While a section of people believed that North Korea was responsible for the death of Otto Warmbier and demanded justice, the others elaborated that it was his own health issues that led to this unfortunate incident.~\footnote{https://en.wikipedia.org/wiki/Otto\_Warmbier}
        \end{itemize}
We next discuss the data related to the news articles and the tweets of each of these events. 
\begin{table}[ht]
\centering
\begin{tabular}{|c|c|c|c|c|c|}
      \hline
      \textbf{Event}  & \textbf{$N_a$} & \textbf{Avg $A_{len}$} & \textbf{Event}  & \textbf{$N_a$} & \textbf{Avg $A_{len}$}\\
      \hline
      {$E_1$} & 34 &  18 & {$E_6$} & 21 &  27\\ 
      \hline
      {$E_2$}  & 86 & 20 & {$E_7$} & 25 &  17\\
      \hline
      {$E_3$}  & 47  & 21 & {$E_8$} & 18 &  35\\
      \hline
      {$E_4$}  & 10  & 32 & {$E_9$} & 32 &  23\\
      \hline
      {$E_5$} & 16  & 18 & {$E_{10}$} & 13 &  45\\
      \hline
\end{tabular}
\caption{The number of news articles, $N_a$  and the average length of a news article, $A_{len}$ related to $10$ events is shown.}\label{tab:newsArtData}
\end{table}

\subsubsection{News Articles and Tweets Dataset}
In this section, we discuss the details related to the news articles of an event and the tweets considered for experimental analysis.\\
\textbf{News Articles Dataset:}
As already discussed in Section~\ref{s:dataset}, we consider different events based on the number of the news articles published related to that event. The total number of news articles published related to an event was considered based on the news articles shown by \textit{Google News Search API} related to the event. We select the news articles related to the events mentioned in Section~\ref{s:eve}. We used \textit{newspaper API} to crawl the randomly selected news articles related to an event. In Table~\ref{tab:newsArtData}, we further provide an overview of the dataset related to the news articles of the events, i.e. \textit{number of news articles} and \textit{average number of sentences in the news article}. Observations from Table~\ref{tab:newsArtData} indicates that each of these characteristics has a high variance, i.e. the \textit{number of news articles} ranges from $10-86$ and \textit{average number of sentences in the news article} ranges from $17-45$. \\
\textbf{Tweets Dataset}~\label{s:twd}
For each of the events, we followed a pseudo relevance feedback based system~\cite{chakraborty2019predicting,chakraborty2017network} to extract the relevant tweets  related to the event and we randomly select $500$ relevant tweets for each event for our experimental analysis. Since we determine the stance of a tweet towards an event from the tweet text, we perform basic pre-processing on the tweet text, like removal of the hashtags, URL and user-ids from the tweet text.


\subsection{Ground Truth Preparation}
We prepare the ground truth data through manual annotation of the stance of the tweets identified in context to each of the events. Manual annotation of the tweets required a contextual understanding of the event to correctly identify the stance of the tweet. As stated above for each of the events we selected $500$ tweets leading to a total of $5000$ tweets. A group of $3$ manual annotators with good knowledge in English and the background of the events were given the task of labelling. An annotator marked the tweet as \textit{positive} or \textit{ negative} with respect to the event. We strongly encouraged them to try their best to understand the tweets before labeling as tweets are sometimes confusing with informal expressions and sarcasm. The  inter-rater  agreement  was  measured using Cohen's kappa coefficient\footnote{https://en.wikipedia.org/wiki/Cohen\%27s_kappa}. We observed a value of $89\%$, thus  indicating  a high  agreement  among  the  annotators. During disagreement, the opinion of the majority of the three annotators was considered.

\subsection{Comparison with Existing Techniques}\label{s:baseline}
We have considered the proven state-of-the-art stance detection approaches as baselines. Several of these approaches are from the \textit{SemEval 2016 Task}~\cite{mohammad2016semeval}.  Although the existing research works (referred by $Base_1$, $Base_2$, $Base_3$ and $Base_4$) is proposed to train the respective models on annotated tweets, we intend to investigate the performance of the \textit{baselines} when trained on \textit{news articles} related to an event. The reason being that as the proposed approach utilizes information extracted from news articles related to an event to determine stance of a tweet, we wanted to investigate the performance of the existing research works when trained on news articles statements rather than tweets. Hence, we randomly selected the number of \textit{news article} statements equal to the number of \textit{tweets} required for training for each of the \textit{baselines} respectively and train the corresponding baselines on news article statements. We refer to these baselines by $Base_5$, $Base_6$, $Base_7$ and $Base_8$ respectively. We briefly discuss all the these baselines next.
\begin{itemize}
\item {\textit{$Base_1$}} : Zarrella et al.~\cite{zarrella2016mitre} used recurrent neural  network  initialized  with  features learned from distant supervision to detect stance in tweets which is referred by $Base_1$ in this paper hereby.

\item {\textit{$Base_2$}} : The work proposed in~\cite{wei2016pkudblab} used a convolutional neural network for stance detection in tweets which would be referred by $Base_2$ in this paper.

\item {\textit{$Base_3$}} : Mohammad et al.~\cite{mohammad2017stance} proposed an ensemble based supervised approach that incorporates an extensive set of content features to detect stance in tweets. This work is referred by $Base_3$ in this paper.

\item {\textit{$Base_4$}} : Tutek et al.~\cite{tutek2016takelab} proposed an ensemble of
learning algorithms and further, fine-tuned using a genetic algorithm. They considered an exhaustive list of features, like word features (unigrams, brigrams), word embeddings, counting features (average word length, number of retweet symbols, number of hashtags, number of emoticons, etc), repeated vowels and  hashtag information. This work is referred by $Base_4$ in this paper.
\item {\textit{$Base_5$}} : Although $Base_1$ was trained on a set of tweets, we trained the same approach on news article statements to check the effectiveness of the approach when trained on news articles. We refer to this baseline as $Base_5$ in this paper.

\item {\textit{$Base_6$}} : We train $Base_2$ on news article statements related to each event rather than tweets and refer it by $Base_6$ in this paper henceforth.

\item {\textit{$Base_7$}} : We refer to $Base_3$ when trained on news article statements related to an event rather than tweets by $Base_7$ in this paper.


\item {\textit{$Base_8$}} :  The baseline $Base_4$ when trained on news articles statements rather than tweets is referred by $Base_8$.
\end{itemize}
We next describe the experiment details and the performance measures that we use for our investigations.

\subsection{Performance Evaluation and Measures}
We rigorously validate the efficiency of the proposed approach. Initially we compare the \textit{accuracy} and \textit{F1} scores with the existing research works. We also compare the effectiveness of the proposed approach when these existing research works are trained on news articles (same dataset as proposed approach) rather than tweets to understand whether the effectiveness of the existing research works would have increased if they were trained on news article statements rather than tweets. Subsequently, we study the properties of the news events to investigate whether the proposed approach is biased towards any specific characteristics. We further investigate the relevance of the target set (both key-phrases and key-players) that we extract using our approach, with respect to the event. Subsequently, we validate the importance of the signed network approach in determining the polarity of the targets towards the event by comparing the efficiency of the \textit{Pass-I} and \textit{Pass-II} steps.
\section{Results and Discussion}\label{s:results}
In this section, we discuss the findings of our experimental investigations. Initially, we compare the performance of the proposed approach with the baseline approaches.

\subsection{Comparison with Baselines}\label{s:res}
For each of the $10$ different events we initially compare the proposed approach with the baselines, in terms of average F1 scores. The results for the same is shown in table~\ref{tab:stanceRes_1}. The proposed approach outperforms the other existing research works by around by $6.5-43\%$ in terms of the average F1 score. The results indicate that the proposed approach is highly efficient in determining the stance of the tweets.

\begin{table*}[ht]
\centering
\begin{tabular}{c|c|c|c|c|c|c|c||c|c|c|c|c|c|}
\multicolumn{2}{c}{}&\multicolumn{2}{c}{{\bf F1-score}}&\multicolumn{2}{c}{}&\multicolumn{2}{c}{}\\
\cline{3-11}
\multicolumn{2}{|c|}{}&Proposed&$Base_{1}$&$Base_{2}$&$Base_{3}$&$Base_{4}$&$Base_{5}$&$Base_{6}$&$Base_{7}$&$Base_{8}$\\
\cline{2-11}
\multirow{2}{*}& $E_{1}$ & $0.80$ & $0.62$& $0.72$ & $0.71$ & $0.69$ &$0.53$&$0.63$&$0.45$&$0.54$\\
\cline{2-11}
\multirow{2}{*}{{}}& $E_{2}$ & $0.61$ & $0.50$& $0.57$ & $0.49$ & $0.50$ &$0.43$&$0.41$&$0.46$&$0.40$\\
\cline{2-11}
\multirow{2}{*}{{}}& $E_{3}$ & $0.63$ & $0.45$& $0.57$ & $0.47$ & $0.46$ &$0.40$&$0.52$&$0.41$&$0.40$\\
\cline{2-11}
\multirow{2}{*}{{}}& $E_{4}$ & $0.66$ & $0.49$& $0.57$ & $0.58$ & $0.50$ &$0.43$&$0.52$&$0.53$&$0.41$\\
\cline{2-11}
\multirow{2}{*}{{}}& $E_{5}$ & $0.70$ & $0.63$& $0.65$ & $0.60$ & $0.63$ &$0.59$&$0.60$&$0.55$&$0.57$\\
\cline{2-11}
\multirow{2}{*}{{}}& $E_{6}$ & $0.69$ & $0.60$& $0.37$ & $0.56$ & $0.54$ &$0.53$&$0.35$&$0.55$&$0.44$\\
\cline{2-11}
\multirow{2}{*}{{}}& $E_{7}$ & $0.70$ & $0.64$& $0.62$ & $0.57$ & $0.62$ &$0.54$&$0.55$&$0.54$&$0.53$\\
\cline{2-11}
\multirow{2}{*}{{}}& $E_{8}$ & $0.76$ & $0.69$& $0.71$ & $0.62$ & $0.69$ &$0.54$&$0.56$&$0.47$&$0.54$\\
\cline{2-11}
\multirow{2}{*}{{}}& $E_{9}$ & $0.80$ & $0.61$& $0.60$ & $0.72$ & $0.74$ &$0.54$&$0.52$&$0.66$&$0.63$ \\
\cline{2-11}
\multirow{2}{*}{{}}& $E_{10}$ & $0.71$ & $0.54$& $0.58$ & $0.57$ & $0.70$ &$0.46$&$0.53$&$0.54$&$0.61$\\
\cline{2-11}
\end{tabular}
\caption{The average F1 score of the proposed approach for different events in comparison to other baselines is shown.}\label{tab:stanceRes_1}
\end{table*}
To obtain an alternative view of the above results  we created a set of $2500$ tweets selected randomly from the $10$ events (details given in section~\ref{s:dataset}). We repeated the experiment on $10$ such different sets, the observations of which are shown in table~\ref{tab:twStanceRes}. We find that the proposed approach ensures an accuracy of around $85-90\%$ and average F1-score of $0.77-0.82$, indicating a high efficiency of the proposed approach. We also observe  that the \textit{baselines} have higher performance when trained on tweets rather than news article statements. The underlying reason probably being the features used in \textit{baselines} are more suitable for \textit{tweets} rather than \textit{news articles} and the text characteristics of the \textit{tweets} are not well captured through these \textit{news articles} texts. We next investigate whether  certain specific properties of the news articles is playing a role in the better performance of the proposed approach.

\begin{table*}[ht]
\centering
\begin{tabular}{|c|c|c|c|c|c|c|c|}
      \hline
      \textbf{Tweet Set} & \textbf{Accuracy} & \textbf{F1-score} & \textbf{Tweet Set} & \textbf{Accuracy} & \textbf{F1-score}\\
      \hline
      {$I_{1}$} & 0.86 &  0.78 & {$I_{6}$} & 0.90  & 0.81\\ 
      \hline
      {$I_{2}$}   & 0.86 & 0.79 & {$I_{7}$} & 0.89 &  0.81\\
      \hline
      {$I_{3}$} & 0.89  & 0.80 & {$I_{8}$}   & 0.87 & 0.80\\
      \hline
      {$I_{4}$} & 0.90 &  0.82 & {$I_{9}$} & 0.86  & 0.77\\ \hline
      {$I_{5}$}   & 0.90 & 0.82 & {$I_{10}$} & 0.85  & 0.77\\
      \hline
\end{tabular}
\caption{The accuracy and average F1 scores of $10$ sets of $2500$ tweets each, selected randomly from the tweets of $10$ events.}\label{tab:twStanceRes}
\end{table*}

\begin{table*}[ht]
\centering
\begin{tabular}{|c|c|c|c|c|c|c|c|c|c|c|}
      \hline
      \textbf{Event} & \textbf{$A(K_l)$} & \textbf{$Munq(K_l)$} & \textbf{$Sunq(K_l)$} & \textbf{$A(K_p)$} & \textbf{$Munq(K_p)$} & \textbf{$Sunq(K_p)$} & \textbf{{$F_s$}} & \textbf{{$F_{cd}$}} & \textbf{{$F_{cx}$}}\\
      \hline
      
      {$E_{1}$} & 14.96 &  5.82 & 9.74 & 5.66 &  2.67 & 1.15 & 0.56 &  0.26 & 0.18\\ 
      \hline
      {$E_{2}$}   & 7.11 & 2.98 & 2.69 & 12.66 & 1.71 & 3.14 & 0.37 & 0.35 & 0.28\\
      \hline
      {$E_{3}$} & 7.59  & 4.00 & 4.77 & 9.61  & 1.55 & 2.77 & 0.23  & 0.44 & 0.33\\
      \hline
      {$E_{4}$}   & 13.18 & 5.46 & 4.16 & 8.72 & 1.81 & 3.92 & 0.38 & 0.35 & 0.20\\
      \hline
      {$E_{5}$} & 10.5  & 4.31 & 3.65 & 14.2  & 3.5 & 4.65 & 0.27  & 0.44 & 0.29\\
      \hline
      {$E_{6}$} & 10.1  & 6.38 & 4.47 & 17.94  & 2.78 & 5.07 & 0.36  & 0.32 & 0.32\\
      \hline
      {$E_{7}$} & 6.72 &  3.86 & 3.87 & 6 &  3.86 & 3.87 & 0.61 & 0.21 & 0.18\\ 
      \hline
      {$E_{8}$}   & 12.05 & 5.0 & 4.69 & 23.05 & 4.22 & 7.14 & 0.77  & 0.076 & 0.16\\
      \hline
      {$E_{9}$} & 6.93  & 4.46 & 4.71 & 6.56  & 1.36 & 2.71 & 0.65  & 0.21 & 0.141\\
      \hline
      {$E_{10}$}   & 16.96 &  8.76 & 18.67 & 10.76 &  2.26 & 8.57 & 0.70  & 0.159 & 0.14\\
      \hline
      
\end{tabular}
\caption{The average number of key-players($A(K_l)$) and key-phrases($A(K_p)$), the average number of unique key-players($Munq(K_l)$) and key-phrases($Munq(K_p)$ and the standard deviation in finding unique key-players($Sunq(K_l)$ and key-phrases($Sunq(K_p)$, fraction of \textit{simple sentence} $F_{s}$, \textit{compound sentence}, $F_{cd}$ and \textit{complex sentence}, $F_{cx}$ in the news articles related to each of the $10$ events is shown }\label{tab:artData_1}
\end{table*}

\subsection{Investigating the News Articles' Dataset}\label{s:Nres}
 We investigate whether the performance of the proposed approach is biased by certain specific properties of the news articles. In this Section, we study characteristics of the news articles and validate whether the differences affects the performance of the proposed model. A high variance in these properties would ensure that the performance of the proposed approach is not dependent on the choice and nature of the articles. We have already highlighted earlier in Section~\ref{s:dataset}, that the news articles selected for the $10$ events exhibit considerable difference in their number and the average length (Table~\ref{tab:newsArtData}). In this Section, we focus on the total count of the \textit{key-players} and \textit{key-phrases} that occurs in the articles and fraction of each type of sentences in the news articles.
 
\par  Our observations as shown in Table~\ref{tab:artData_1} indicate that \textit{average number of key-players} in the articles range from $6.72$-$16.96$, the \textit{average number of unique key-players} in the articles range from $2.98$-$8.76$ and the \textit{standard deviation in number of unique key-players} of an article ranges from $2.67$-$18.67$, the \textit{average number of key-phrases} in the articles range from $5.66$-$23.05$, the \textit{average number of unique key-phrases} in the articles range from $1.55$-$4.22$ and the \textit{standard deviation in number of unique key-phrases} of an article ranges from $1.15$-$8.57$. On investigating the distribution of different types of sentences in the news articles related to each of the events, we observe as shown in Table~\ref{tab:artData_1} that there is considerable variance. For example,  \textit{Harvey Weinstein Allegations}, \textit{Australian Ball
Tampering Scandal} and \textit{Assasination of Jamal Khassogi} have similar distribution of \textit{simple} and \textit{complex} sentence while  \textit{Demonetization in India} has a majority of \textit{simple sentence} and \textit{Catalan Independence Movement} and \textit{Punjab National Bank Scam} have a majority of \textit{complex sentences}. Thus, these results shows that the considerably high performance of the proposed approach for all the $10$ events is unlikely due to any specific characteristics of the news articles and the proposed approach is not biased by the type of news articles or event.

\par
\begin{figure*}[ht]
\centerline{
\subfigure[$E_1$]{\label{fig:h}\includegraphics[width=2in]{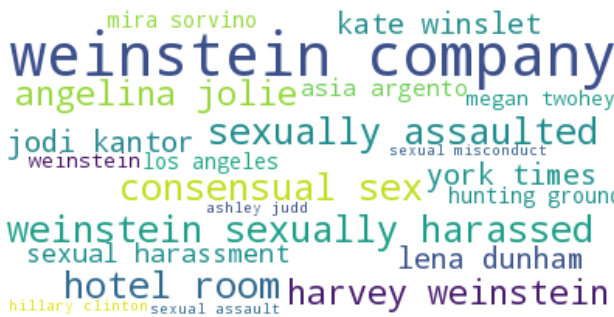}}
\subfigure[$E_2$]{\label{fig:d}\includegraphics[width=2in]{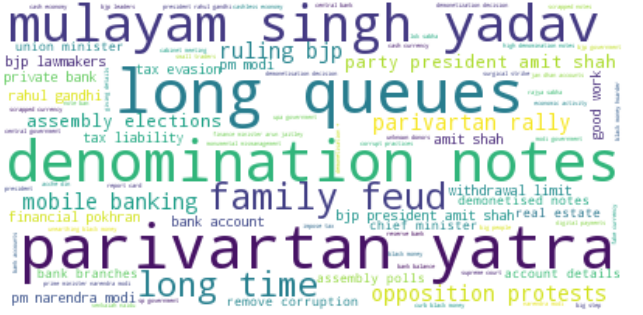}}
\subfigure[$E_4$]{\label{fig:c}\includegraphics[width=2in]{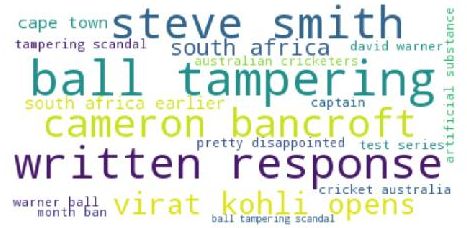}}
}

\caption{Word cloud representing the \textit{key-phrases} generated related to each of the $3$ events is shown.}
\label{fig:rake_1}
\end{figure*}

 Further, we highlight on the target \textit{key-phrases} that are extracted using our approach by representing a sub-set of the identified \textit{key-phrases} using a word cloud related to $3$ events in Figure~\ref{fig:rake_1}. For example, for the event, namely \textit{Harvey Weinstein allegations}, prominent personalities like Angelina Jolie, Kate Winslet, Lena Dunham and Jodi Kantor who made allegations against him get strongly represented. Further major key-phrases like \textit{sexual harrassment, sexual assault} and \textit{consensual sex} that were key features of the debate are also suitably captured. Similarly, for the demonetization event in India, key-phrases like \textit{long queues}, lack of certain \textit{denomination notes} and \textit{parivartan} (change) that were being highly used to debate on the benefits and drawbacks of demonetization are captured successfully.

\subsection{Investigating the Polarity of Targets towards the Event}\label{s:tPol}

In this Section, we discuss the effectiveness of the proposed approach in determining the polarity relationship of a target towards the event through \textit{Pass-I} and \textit{Pass-II} respectively. Based on our earlier investigations in Section~\ref{s:sgnW}, we observed that the fraction of targets related to an event whose polarity could be extracted in \textit{Pass-I} ranged from $14-62.5\%$, in \textit{Pass-II} ranged from $31.3-76\%$ (table~\ref{tab:missLink}). On studying the correctness of the polarity relationships extracted in both \textit{Pass-I} and \textit{Pass-II} respectively, we observe that the proposed approach achieves $100\%$ correctness in determining the  polarity of a target in both \textit{Pass-I} and \textit{Pass-II}.  
\par However, even though the proposed approach can correctly identify the polarity of the targets in  both \textit{Pass-I} and \textit{Pass-II}, there remains a small fraction of targets(around $4-12\%$) whose polarity relationship with the event could not be extracted in both \textit{Pass-I} and \textit{Pass-II}. The polarity relationship of a target, say $g_{i}$ towards the event, $E$, could not be resolved in \textit{Pass-I} in any of the following 2 cases: ($a$) if the target appeared in only news article statements that are direct or indirect speech in which the \textit{reporting clause} comprised of complex or compound sentences or ($b$) if the target appears in complex and compound sentences where the conjunction present is not one of the conjunctions as shown in Table~\ref{tab:ConjData}. Further, the polarity relationship of $g_{i}$ towards the $E$ would not be resolved even in \textit{Pass-II} if it is not connected to  any other target $g_j$ which is connected to $E$ through some path. Therefore, the proposed approach fails to extract the polarity relationship of $g_{i}$ towards $E$ if it satisfies both the above mentioned conditions for \textit{Pass-I} and \textit{Pass-II}, respectively. However, if neither of these conditions hold true for $g_{i}$, then the proposed approach can extract the polarity relationship of $g_i$ towards $E$ with $100\%$ accuracy. 


\subsection{The role of sentence types in identifying the polarity of a target towards the event}\label{s:simPOl}
extract the polarity relationship of the targets towards an event (\textit{Pass-I}). While  extracting the polarities from the simple sentences is simpler and precise as compared to the other two types, however, the fraction of simple sentences in the news articles may be very less. 
In this Section, we investigate the loss in efficiency of the proposed approach had we considered only simple sentences from the news articles, i.e., we study the requirement of all the three types of sentences than only simple sentences. 
We observe the fraction of targets whose polarity relationship towards the event could be extracted by handling only \textit{simple} sentences and \textit{direct and indirect} speeches where the reporting clause is a \textit{simple} sentence in Table~\ref{tab:newsArtSim}. Our observations indicate that using all the three types of sentences for target polarity identification increases the efficiency by more than $40\%$. 

\begin{table}[ht]
\centering
\begin{tabular}{|c|c|c|c|c|c|c|}
      \hline
      \textbf{Sentence} & \textbf{$E_{1}$} & \textbf{$E_{2}$} & \textbf{$E_{3}$} & \textbf{$E_{7}$} & \textbf{$E_{8}$}\\
      \hline
      {$Simple$} & $0.37$ &  $0.10$ & $0.11$ & $0.20$ & $0.238$\\ 
      \hline
      {$Pass-I$}   & $0.527$ &  $0.14$ & $0.277$ & $0.303$ & $0.351$ \\
      \hline
\end{tabular}
\caption{The fraction of \textit{targets} whose polarity towards an event was extracted by considering only \textit{simple} sentences and \textit{direct and indirect} speech where the reporting clause is \textit{simple} (represented by $Simple$) and by following \textit{Pass-I} is shown. The number of events considered is $5$.}\label{tab:newsArtSim}
\end{table}

\subsection{Analyzing failures}
\label{s:fail}
Although the proposed approach ensures high accuracy in comparison with the existing baseline approaches irrespective of the type of event, we investigate more closely certain cases where our proposed approach fails. 
\par \textit{Extraction of Polarity Relationship from News Article} : A typical example sentence (related to the Demonetization event) where the proposed approach fails to identify the polarity of the target towards the event is, \textit{The opposition party leader claimed ``Only $6\%$ of the black money is in cash form while $94\%$ of the black money is either in foreign banks or in the form of gold, land and estate''}. A major characteristic of this sentence is it contains a
latent domain information related to the event. The sentence provides an information that black money in cash is much less as compared to other forms; however, it subtly hints that since demonetization would mainly impact black money in cash, hence it would be unsuccessful. Thus, the polarity relation of the target \textit{black money in cash} with the demonetization event would not be resolved and hence the stance of this tweet cannot be determined by our proposed approach. 
\par \textit{Identifying Target Polarity using Signed Network Analysis} :  Our proposed approach fails to predict the polarity of a target towards the event if the target (say, $a$) is connected to only another target (say, $b$) and neither $a$ nor $b$ is connected to any other target or the event in the signed network. Therefore, the proposed approach fails to extract the polarity relationship of a target towards the event in \textit{Pass-II} if it does not satisfy either of the conditions mentioned in Section~\ref{s:missL}. Our empirical experiments indicate the fraction of the targets whose polarity could not be resolved by the proposed methodology ranges from $4-12\%$ of the total targets for the $10$ events (as shown in Table~\ref{tab:missLink}). Thus, there remains a scope of improvement of the proposed approach so as to resolve these cases. 
 \par \textit{Stance Detection of a Tweet towards the Event} :  There are few scenarios in which the proposed approach in Section~\ref{s:senTw} fails to determine the stance of a tweet towards an event. One of them being  processing of \textit{sarcasm} if present in tweets. Understanding of \textit{sarcasm} in tweets requires in-depth understanding of both the content and context level information in the tweet text~\cite{joshi2017automatic} which has not been covered in the current proposed approach. The proposed approach also fails to extract stance of a tweet if the tweet text is written in any other language except \textit{English}. Although the proposed approach follows an exhaustive mechanism to extract the targets related to an event and we have experimentally verified the success of this step, it cannot ensure that all the related targets of an event would be captured. Therefore, if a tweet comprises of a target that has not been identified previously by the proposed approach, it fails to identify the stance of the tweet towards the event. Therefore, these issues provide directions where the proposed approach can be further improved.
 \section{Conclusion}\label{s:concl}
In this paper, we propose an unsupervised approach to detect stance in tweets. The novelty of the proposed approach is the use of a signed network to detect stance of a tweet. Though signed networks have been proposed to understand the relationships between the actors of a text/movie review, this is the first work that leverages the information from the signed network for stance detection in tweets. The created signed network serves as a rich knowledge base related to an event --- comprising of the possible exhaustive target set related to the event and polarity relationships between any pair of targets which is extracted by a combination of textual and signed network attributes. This further ensures that stance detection by the proposed approach does not require any human intervention. Validation on empirical data set of around $10$ events shows that stance detected using the proposed approach outperforms the existing stance detection approaches by a considerable margin.

\bibliographystyle{acm}
\bibliography{stance.bib}
\end{document}